\documentstyle[aps,epsf,amssymb,amscd]{revtex}
 \newcommand{\eins}{\mbox{$1 \hspace{-1.0mm}  {\bf l}$}}
 
 \newcommand{\be}{\begin{equation}}
 \newcommand{\ee}{\end{equation}}
 \newcommand{\bea}{\begin{eqnarray}}
 \newcommand{\eea}{\end{eqnarray}}

 \newcommand{\ket}[1]{ | \, #1  \rangle}
 \newcommand{\bra}[1]{ \langle #1 \,  |}
 \newcommand{\proj}[1]{\ket{#1}\bra{#1}}

\newcommand{\kett}[1]{ |  #1  \rangle\!\rangle}
\newcommand{\braa}[1]{ \langle \!\langle #1   |}
\newcommand{\scall}[2]{ \langle \!\langle #1   | #2
\rangle\!\rangle}

\newcommand{\projc}[1]{\ket{#1}_c \,{}_c\bra{#1}}
\newcommand{\proja}[1]{\ket{#1}_a \,{}_a\bra{#1}}

\begin{document}
\twocolumn
                 \title{Joint measurements via quantum cloning}
                 \author{G. M. D'Ariano $^{a}$, C. Macchiavello $^{a}$ and
		 M. F. Sacchi $^{a,b}$}
                 \address{${}^a$ Theoretical Quantum Optics Group\\ 
                 Dipartimento di Fisica ``A. Volta'' and INFM - Unit\`a 
                 di Pavia,
                 Via A. Bassi 6, 27100 Pavia, Italy\\${}^b$ 
                 Optics Section, The Blackett Laboratory, Imperial College, London SW7 2BW, United Kingdom}
                 \date{Received \today}
                 \maketitle
                  \begin{abstract}

We explore the possibility of achieving optimal joint measurements of
noncommuting observables on a single quantum system by performing
conventional measurements of commuting self adjoint operators on
optimal clones of the original quantum system. We consider the case of
both finite dimensional and infinite dimensional Hilbert spaces. In
the former we study the joint measurement of three orthogonal
components of a spin 1/2, in the latter we consider the case of the
joint measurements of any pair of noncommuting quadratures of one
mode of the electromagnetic field. We show that universally covariant
cloning is not ideal for joint measurements, and a suitable non universally 
covariant cloning is needed.

                 \end{abstract}
                 \pacs{03.67.-a, 03.65.-w}
                 %\twocolumn
                 %\widetext

\section{Introduction}

The first scheme for the  joint measurement of noncommuting observables
performed on a single quantum system was introduced by Arthurs-Kelly
\cite{A-K}. The problem of evaluating the minimum added noise in the
joint measurement of position and momentum, and more generally of a
pair of observables whose commutator is not a c-number was then solved
by Yuen \cite{Yu}.  A similar approach to the problem has been
followed in Ref. \cite{goodman}.  In the case of two quadratures of
one mode of the electromagnetic field the problem can be phrased in
terms of a coherent POVM whose Naimark extension introduces an
additional mode of the field. This kind of measurement can be realised
by means of a heterodyne detector \cite{heterodyne}.

The case of the angular momentum of a quantum system is more difficult, 
and no measurement scheme has appeared in the literature
so far. Spin coherent states \cite{arecper}
can be introduced and interpreted as continuous (overcomplete) POVM, 
but the corresponding Naimark extension is unknown. It
was shown that the spin coherent POVM minimises suitably defined
quantities that represent the precision and the disturbance of the
measurement \cite{Appleby}, but explicit realisations of such a POVM
are not known \cite{peres,lopresti}.  The joint measurement of the
three components $J_x$, $J_y$ and $J_z$ of the angular momentum could
also be studied with discrete spectrum, rather than continuous. 
This problem does not have a solution
yet. Joint measurements are a crucial ingredient in general quantum
teleportation schemes \cite{telep}, and are essential in connecting the
quantum with the classical meaning of the angular momentum
itself. Therefore, it is of great interest to find schemes that
realise them.

The idea of this paper is to use quantum cloning to achieve joint 
measurements. It is well known that perfect cloning of unknown quantum
systems is forbidden by the laws of quantum mechanics \cite{wootters}.
The first universal cloning machine for spin 1/2 systems was proposed in 
Ref. \cite{bh}, and later proved to be optimal in \cite{oxibm}. 
More general universal transformations were then proposed in \cite{gm} and 
proved to be optimal in Refs. \cite{bc,Wer} (in Ref. \cite{Wer} the CP map 
of the optimal cloning transformations for finite dimensional systems was 
derived). However, the complete unitary transformation achieving the optimal cloning 
is not known (in Ref. \cite{gm} some matrix elements of the unitary transformation 
are given for the case of qubits). 

If we want to use quantum cloning to realise joint measurements, we may
need to optimise it for a reduced covariance group, depending on the
kind of the desired joint measurement. The cloning transformations
mentioned above were optimised by imposing total covariance, i.e. for
all possible unitary transformations. In general a restriction of the covariance
group  leads to a higher fidelity of the cloning
transformation, as for example in the case of phase covariant cloning
\cite{pcc}, where,  however,  only the bounds for the fidelity of
the optimal cloning are given, but not the form of the optimal map.

For infinite dimensional systems it is not clear how to find the
universal transformations for cloning. The extension to infinite dimension 
of the maps given in Ref. \cite{Wer} needs a regularization procedure, an example of
which is given here in section \ref{reg}. The infinite dimensional $1\to
2$ cloning machine proposed in \cite{cerf} is universal for coherent
states, with resulting fidelity equal to $2/3$.  In this paper we show
that the cloning transformations proposed in \cite{cerf} are optimal
for the joint measurement of orthogonal quadratures, and the joint
measurement can be generalised to any angle between two noncommuting
quadratures.

In the case of finite dimensional systems we will study the joint
measurement of the three components of spin 1/2 states by operating a
$1\to 3$ universal optimal cloning transformation on the original
state and then performing independent measurements of $\sigma_x$,
$\sigma_y$ and $\sigma_z$ on the three output copies.  We will show
that the resulting POVM is not optimal with respect to the
added noise, but it gives a pretty good approximation of the optimal one.

The paper is organised as follows. In section II we consider the case
of spin 1/2 systems. We first recall the optimal universal $1\to 3$
cloning transformations and then exploit them to achieve joint spin
measurements.  In section III we study the case of infinite
dimensional systems, first reviewing the optimal $1\to 2$
transformation of Ref. \cite{cerf} and then applying it to the joint measurement of two
quadratures of one mode of the electromagnetic field.  In section IV
we present a regularization of the map in Ref. \cite{Wer} in order to
extend it to infinite dimensional Hilbert spaces, and show that the
universal cloning does not achieve the optimal joint
measurement. We summarise the results in section IV.

\section{The finite dimensional case: joint spin measurements}\label{fin}

In this section we analyse the case of spin 1/2 systems, by first reviewing
the optimal universal $1\to 3$ cloning transformations which produce three
output copies from a single input, and then exploiting this procedure to 
achieve joint measurements of the spin components. We will show that the 
joint spin measurement obtained in this way is only an approximation of
 the spin measurement POVM of Ref. \cite{arecper}.

\subsection{Optimal $1\to 3$ cloning}

We consider the case of universal cloning, namely transformations whose
efficiency does not depend on the form of the input state.
General $N\to M$ universal cloning transformations, which act on $N$ copies
of a pure state $\ket{\psi}$ 
and produce $M$ output copies as close as possible to the
input state, were proposed in Ref. \cite{gm} and later proved to be 
optimal in Refs. \cite{bc,Wer}.
We consider here the form given in Ref. \cite{Wer}. The output state 
$\rho_M$ of the $M$ copies for spin 1/2 systems is given by
\begin{eqnarray}
\rho_M=\frac{N+1}{M+1}S_M\left(\ket{\psi}\bra{\psi}^{\otimes N}\otimes 
\eins^{\otimes(M-N)}\right)S_M\;,
\label{tr-wer}
\end{eqnarray}
where $S_M$ is the the projection operator onto the symmetric subspace of 
the $M$ output copies.
The fidelity $F(N,M)=\bra{\psi}{\mbox{Tr}}_{M-1}[\rho_M]\ket{\psi}$
of each output copy with respect to the initial state
$\ket{\psi}$ is given by
\begin{eqnarray}
F(N,M)=\frac{M(N+1)+N}{M(N+2)}\;.
\label{F-NM}
\end{eqnarray}
Since the cloning transformation is universal it can be also viewed as a
shrinking transformation of the Bloch vector of each copy, described 
by the shrinking factor $\eta(N,M)$ \cite{oxibm,bc}: 
the density operator describing the state of the $M$ output
copies is given by
$\rho_{out}=\frac{1}{2}[\eins+\eta(N,M)\vec{s}_{in}\cdot \vec{\sigma}]$,
where $\vec{s}_{in}$ denotes the Bloch vector of the initial state
$\ket{\psi}$ and $\{\sigma_\alpha ,\alpha =x,y,z\}$ are the Pauli operators. 
For the optimal transformations (\ref{tr-wer}) the shrinking
factor is 
\begin{eqnarray}
\eta(N,M)=\frac{N}{M}\frac{M+2}{N+2}\;.
\label{eta-NM}
\end{eqnarray}
In the particular case  of the $1\to 3$ cloning transformation, which we
will consider in the following, the above map takes the form
\begin{eqnarray}
\rho_3=\frac{1}{2}S_3\left(\ket{\psi}\bra{\psi}\otimes 
\eins^{\otimes 2}\right)S_3\;,
\label{tr-3}
\end{eqnarray}
where $S_3$ is the projector on the space spanned by the vectors 
$\{\ket{s_i}\bra{s_i}, i=0\div3\}$,
with $\ket{s_0}=\ket{000}$, $\ket{s_1}=1/\sqrt{3}(\ket{001}+
\ket{010}+\ket{100})$, $\ket{s_2}=1/\sqrt{3}(\ket{011}+
\ket{101}+\ket{110})$ and $\ket{s_3}=\ket{111}$, where $\{\ket{0},\ket{1}\}$
is a basis for each spin 1/2 system. The value of the shrinking factor in
this case is
$\eta(1,3)=5/9$.

\subsection{The joint spin measurement via cloning}

We will now study a method to measure jointly the three components of a 
spin 1/2 system by first generating three approximate copies of the 
input state through an optimal $1\to 3$ cloning transformation, and then 
performing independent measurements on the three copies, namely measuring 
a different spin component on each copy. The POVM corresponding to the usual
projection measurement of the $\alpha $-component of the Bloch vector on one copy
is given by the operator $[\eins+m_\alpha \sigma_\alpha ]/2$, where $\alpha =x,y,z$ and 
$m_\alpha = \pm 1$ corresponds to the outcome of the measurement.
The POVM $\Omega (\vec m)$ describing the measurement of the three components, each performed on
a different copy, is then given by
\begin{eqnarray}
\Omega (\vec m)=\frac{1}{8}(\eins + m_x\sigma_x)\otimes (\eins + m_y\sigma_y)\otimes
(\eins + m_z\sigma_z)\;,
\label{pom-3}
\end{eqnarray}
where the triplet $\{m_x,m_y,m_z\}$ represents the outcomes of the measurement. 
We will now consider the sequence of the $1\to 3$ cloning transformation 
followed by the measurement of a spin component on each of the three
copies as a joint measurement on the initial input state of the original 
copy. In order to derive the corresponding POVM we first compute the 
probability distribution $p(\vec m)$ as a function
of the vector $\vec m=\{m_x,m_y,m_z\}$ of the outcomes
\begin{eqnarray}
p(\vec m)&=&{\mbox{Tr}}
\left[\Omega (\vec m)\,
\frac{1}{2}\,S_3
\left(\ket{\psi}\bra{\psi}\otimes \eins^{\otimes 2}\right)S_3\right ]
\nonumber \\&=&{\mbox{Tr}}_{1}\left[\ket{\psi}\bra{\psi}
\frac{1}{2}{\mbox{Tr}}_{2,3}
[S_3\,\Omega (\vec m)\, S_3]\right]\;,
\label{pi3}
\end{eqnarray}
where Tr$_i$ denotes the partial trace over the $i$th clone. 
This measurement, viewed as a  joint measurement on the original copy 
$\ket{\psi}\bra{\psi}$ can then be described in terms of the POVM $\Pi(\vec m)$
\begin{eqnarray}
\Pi(\vec m)&=&
\frac{1}{2}{\mbox{Tr}}_{2,3}
\left[S_3\,\Omega (\vec m)\,S_3\right]\;.
\label{povm}
\end{eqnarray}
A lengthy and straightforward matrix algebra gives  the resulting POVM in the simple form
\begin{eqnarray}
\Pi(\vec m)=\frac{1}{8}\,\left[\eins+\frac{5}{9}\vec m \cdot\vec\sigma\right]\;.
\label{povms}
\end{eqnarray}
Notice that the $5/9$ factor in front of the Pauli operators corresponds to the
shrinking factor of the optimal $1\to 3$ cloning transformation.
This can be intuitively 
expected because the average value of the spin components of the 
three cloned copies that are measured is shrunk by this factor.

We will now compute the accuracy of this joint measurement in order to have a 
comparison with the coherent POVM given in Ref. \cite{arecper}. 
As mentioned above, the POVM (\ref{povms}) leads to the following 
rescaling between the measured average value $\langle \sigma_\alpha \rangle_m$ 
and the theoretical one for all
the three spin components
\begin{eqnarray}\langle \sigma_\alpha \rangle_m=\sum_{\vec m} m_\alpha \, {\mbox{Tr}}
[\ket{\psi}\bra{\psi}\,\Pi(\vec m)]
=\frac{5}{9}\,\langle \psi | \sigma_\alpha | \psi \rangle \;.
\label{resc}
\end{eqnarray}
Therefore, the unbiased
estimate $\langle \sigma_\alpha \rangle_e$ for the spin components corresponds
to rescaling the measured outcome variables to
$m_\alpha =\pm9/5$, such that
\begin{eqnarray}
\langle \sigma_\alpha \rangle_e=\frac{9}{5}\,\langle \sigma_\alpha \rangle_m\;
\label{resc2}
\end{eqnarray}
and the second moment is also rescaled as follows 
\begin{eqnarray}
\langle \Delta\sigma_\alpha ^2\rangle_e=\frac{81}{25}\,\langle \Delta
\sigma_\alpha ^2\rangle_m\;.
\label{sec}
\end{eqnarray}
In order to study the uncertainty of this measurement we compute the sum 
of the variances corresponding to the three spin components 
$J_\alpha =\sigma_\alpha /2$. 
Since $\langle \sigma_\alpha ^2\rangle_m=1$ for all the components, the uncertainty
in the estimate is given by 
\begin{eqnarray}
\langle \Delta J^2\rangle_e&=&\sum_{\alpha =x,y,z}\langle J_\alpha ^2\rangle_e
-\langle J_\alpha \rangle_e^2\nonumber\\
&=&\frac{1}{4}\left(3 \frac{81}{25}-1\right)=\frac{109}{50}\;.
\label{J2}
\end{eqnarray}

We will now compute the corresponding accuracy for the coherent measurement 
\cite{arecper}. The coherent POVM is given by the projection onto spin coherent
states $\ket{\bf{n}}\bra{\bf{n}}$ \cite{Perelomov}, 
where ${\bf{n}}=(\sin\theta\cos\phi,\sin\theta\sin\phi,\cos\theta)$ 
is a unit vector and 
\begin{eqnarray}
{\bf{n}}\cdot{\bf{J}}{\ket{\bf{n}}}=-j\ket{\bf{n}}\;.
\label{scs}
\end{eqnarray}
Let us calculate as an example the uncertainty related to the component 
$J_z$. Since the measurement is unbiased, the measured mean values of the spin
components $\langle J_\alpha \rangle_m$ coincide with the theoretical mean values, and we 
don't need to introduce rescaling factors as we did in the previous case.
The estimated values therefore coincide with the measured ones. 
For the component $J_z$ one has \cite{Perelomov}
\begin{eqnarray}
\langle J_z\rangle_m=
\int d\mu({\bf{n}})(j+1)\cos\theta\ket{\bf{n}}\bra{\bf{n}}\;,
\label{jz}
\end{eqnarray}
where $d\mu({\bf{n}})=d{\bf{n}}(2j+1)/4\pi$. The measured
mean value of $J_z^2$ is given by
\begin{eqnarray}
\langle J_z^2\rangle_m=
\int d\mu({\bf{n}})(j+1)^2\cos^2\theta\ket{\bf{n}}\bra{\bf{n}}\;,
\label{j2z}
\end{eqnarray}
that can be written as \cite{Perelomov}
\begin{eqnarray}
\langle J_z^2\rangle_m=
\frac{j+1}{j+3/2}\left[\langle \psi | J_z^2 |\psi \rangle +\frac{1}{2}(j+1)\right]\;.
\label{j2z2}
\end{eqnarray}
The measured mean values related to the $x$ and $y$ components can be 
calculated analogously and one has the same relation as Eq. (\ref{j2z2}) for all 
components $\alpha =x,y,z$. 
The total uncertainty in the spin measurement then takes the form
\begin{eqnarray}
\langle \Delta J^2\rangle_e&=&\frac{j(j+1)^2}{j+3/2}+3\frac{(j+1)^2}{2j+3}
-\sum_{\alpha =x,y,z}\langle J_\alpha \rangle_e^2\nonumber\\
&\geq &2j+1\;,
\label{J2c}
\end{eqnarray}
where for $j=1/2$ and pure states the bound is achieved, and is equal to 2. 
This value has to be compared with Eq. (\ref{J2}), obtained by three 
measurements on the three cloned copies.
As we can see, the joint measurement via cloning does not achieve the minimum added 
noise as the optimal POVM of Eqs. (\ref{scs}) and (\ref{jz}), however it  
provides a good approximation. Notice that the minimum added noise would be achieved 
by a discrete POVM of the form 
$\Pi(\vec m)=\frac{1}{8}[\eins+\vec m \cdot\vec\sigma]$.

\section{The infinite dimensional case: joint quadrature measurements}
\label{infc}
In this section we study the cloning for infinite dimensional systems
proposed in Ref. \cite{cerf}.  We review the optimal $1\to 2$
transformation and then apply it to the joint measurement of two
quadratures of one mode of the electromagnetic field. We will show
that the cloning transformation is optimal for joint measurements of
orthogonal quadratures, and the joint measurement can be generalised
to any angle between two noncommuting quadratures by suitably
changing the state of the ancilla. 
\subsection{Optimal $1\to 2$ cloning}
For the following, it is convenient to introduce the formalism of
heterodyne eigenvectors. Consider the heterodyne-current operator\cite{yuen} 
$Z=a+b^\dag$, which satisfies the commutation relation 
$[Z,Z^\dag]=0$ and the eigenvalue equation $Z
\kett{z}_{ab}=z\kett{z}_{ab}$, with $z\in\mathbb {C}$.  The
eigenstates $\kett{z}_{ab}$ are given by \cite{zeta1,zeta2}
\begin{eqnarray}
\kett{z}_{ab}\equiv D_a(z)\kett{0}_{ab}=D_b(z^*)\kett{0}_{ab}\;, 
\end{eqnarray}
where $D_d(z)=e^{zd^\dag -z^* d}$ denotes the displacement operator
for mode $d$ and $\kett{0}_{ab}\equiv
\frac{1}{\sqrt\pi}\sum_{n=0}^{\infty}(-)^n \ket{n}_a\ket{n}_b$.  The
eigenstates $\kett{z}_{ab}$ are a complete orthogonal set with
Dirac-normalization ${}_{ab}\scall{z}{z'}_{ab}=\delta ^{(2)}(z-z')$,
$\delta ^{(2)}(z)$ denoting the delta function over the complex
plane. For $z=0$ the state $\kett{0}_{ab}$ can be approximated by a
physical (normalizable) state, corresponding to the output of a
non-degenerate optical parametric amplifier (NOPA)---so-called twin
beam---in the limit of infinite gain at the NOPA \cite{zeta1}. \par It
is also useful to evaluate the expression ${}_{cb}\scall{z}{z'}_{ab}$
which is given by
\begin{eqnarray}
{}_{cb}\scall{z}{z'}_{ab}=
\frac 1\pi D_a(z'){\cal T}_{ac}D_c^\dag (z)
\;,\label{usef}
\end{eqnarray}
where ${\cal T}_{ac}=\sum_n |n\rangle_a \,{}_c\langle n| $ denotes the
{\em transfer} operator \cite{telep} satisfying the relation ${\cal
T}_{ac} \ket{\psi}_c= \ket{\psi}_a$ for any state $|\psi \rangle $. In
the following we transpose the main results of the continuous variable
cloning of Ref. \cite{cerf}, according to the formalism just
introduced.  The input state at the cloning machine can be written
\begin{eqnarray}
\ket{\phi}=\ket{\varphi}_c\otimes \int_ {\mathbb C} d^2 z\,f(z,z^*)
\kett{z}_{ab}\;\label{fphi} 
\end{eqnarray}
where $\ket{\varphi}_c$ is the initial state to be cloned, belonging
to the Hilbert space ${\cal H}_c$, whereas ${\cal H}_a$ is the Hilbert
space pertaining to the cloned state, and ${\cal H}_b$ is an ancillary 
Hilbert space.  We do not specify for the moment the
explicit form of the function $f(z,z^*)$.  The cloning transformation
is realized by the unitary operator
\begin{eqnarray}
U&=&
\exp\left[c(a^\dag +b )-c^\dag (a+b^\dag)\right ]\nonumber \\&= & 
\exp\left[2i\left(Y_c \hbox{Re}Z-X_c\hbox{Im}Z\right)\right]\;\label{uk}
\end{eqnarray}
with $X_c,Y_c$ denoting the conjugated quadratures for mode
$c$, namely $X_c=(c+c^\dag )/2$ and $Y_c=(c-c^\dag)/2i$. 
\par The unitary evolution in
Eq. (\ref{uk}) can be approached experimentally by means of a network of three 
NOPA's under suitable gain conditions \cite{dds}. 
Notice the simple relation $U
\kett{z}_{ab}=D_c^{\dag}(z)\,\kett{z}_{ab}$. The state after the
cloning transformation is given by 
\begin{eqnarray}
\ket{\phi _{out}}=U\ket{\phi}=
\int_{\mathbb C}d^2z\,
f(z,z^*)\,D_c^{\dag}(z) \ket{\varphi}_c\otimes 
\kett{z}_{ab}\;.\label{fout}
\end{eqnarray}
Let us evaluate the one-mode restricted density matrix $\varrho_c$ and
$\varrho_a$ corresponding to the state $\ket{\phi_{out}}$, for the
Hilbert spaces ${\cal H}_c$ and ${\cal H}_a$ supporting the two clones.
For $\varrho_c$ one has
\begin{eqnarray}
\varrho_c&=&\hbox{Tr}_{ab}[\proj{{\phi _{out}}}] \nonumber \\&=& 
\int_{\mathbb C}d^2w\int_{\mathbb C}d^2z\int_{\mathbb C}d^2z'\,
f(z,z^*)f^*(z',z'^*)\times \nonumber \\&& 
{}_{ab}\braa{w}D_c^{\dag}(z) \ket{\varphi}_c\otimes 
\kett{z}_{ab}\,\,
{}_c\bra{\varphi}D_c(z')\otimes {}_{ab}{\scall{z'}{w}}_{ab}\nonumber
\\&= & \int_{\mathbb C}d^2z |f(z,z^*)|^2 D_c^{\dag}(z)\projc{\varphi}
D_c(z)\;,\label{cl1}
\end{eqnarray}
where we have evaluated the trace by using the completeness and the
orthogonality relation of the eigenstates $\kett{w}_{ab}$.  For
$\varrho_a$, using Eq. (\ref{usef}), one has
\begin{eqnarray}
\varrho_a&=&\hbox{Tr}_{cb}[\proj{{\phi_{out}}}] \nonumber \\&=&
\int_{\mathbb C}d^2w\int_{\mathbb C}d^2z\int_{\mathbb C}d^2z'\,
f(z,z^*)f^*(z',z'^*)
\times \nonumber \\&& 
{}_{cb}\braa{w}D_c^{\dag}(z) \ket{\varphi}_c\otimes 
\kett{z}_{ab}\,\,
{}_c\bra{\varphi}D_c(z')\otimes {}_{ab}{\scall{z'}{w}}_{cb}
\nonumber \\&= & 
\int_{\mathbb C}d^2w\int_{\mathbb C}\frac{d^2z}{\pi}\int_{\mathbb C}
\frac{d^2z'}{\pi}\, f(z,z^*)f^*(z',z'^*) \nonumber \\&\times & D_a(z
){\cal T}_{ac}\left[ D_c^{\dag}(w)D_c^{\dag}(z)
\projc{\varphi}D_c(z')D_c(w)\right]\nonumber \\&\times & {\cal
T}_{ca}D_a^\dag (z')
\nonumber \\& =& 
\int_{\mathbb C} d^2 w\, | \stackrel {\sim }{f}
(w,w^{*})|^2 D_a^{\dag }(w) \proja{\varphi}D_a(w) \;,\label{after}
\end{eqnarray}
where $\stackrel{\sim }{f}(w,w^{*})$ denotes the Fourier transform
over the complex plane
\begin{eqnarray}
\stackrel{\sim }{f}(w,w^*)=\int_{\mathbb C}
\frac{d^2z}{\pi}\,e^{wz^*-w^*z}\,f(z,z^*)
\;. \end{eqnarray}
Hence, for $f(z,z^*)=\stackrel{\sim }{f}(z,z^*)$ 
one has $\varrho_c=\varrho_a$, namely the two clones are identical.
In the following we will show that the choice of the function
$f(z,z^*)$ determines a criterium of optimality in terms of joint
measurement of noncommuting quadratures of the original system through
the measurement of separate (commuting) quadratures over the two clones. 
\subsection{The joint quadrature measurement via cloning}
Quantum cloning allows one to engineer new joint measurements of a
quantum system, by suitably measuring the cloned copies. In the case
of $1\rightarrow 2$ copies just introduced, measuring two quadratures
on the two clones is equivalent to the joint measurement of conjugated
quadratures on the original, similarly to a heterodyne
measurement. Consider the simplest case
\begin{eqnarray}
f(z,z^*)=\sqrt{\frac 2\pi}\,\exp(-|z|^2)
\;\label{fz}
\end{eqnarray}
 in Eqs. (\ref{fphi}), (\ref{cl1}) and
(\ref{after}). One
obtains $\varrho_c=\varrho_a$, namely the two clones are identical,
and their state is given by the original state $|\varphi \rangle $ degraded by
Gaussian noise. The state preparation $\ket{\chi}$ pertaining to the
Hilbert space ${\cal H}_a\otimes {\cal H}_b$ is given explicitly by
\begin{eqnarray}
\ket{\chi}&=&
\sqrt{\frac 2\pi}\int_{\mathbb C}
d^2z\,e^{-|z|^2} \kett{z}_{ab} \nonumber \\& =&
\sqrt{\frac 2\pi}\int_{\mathbb C}
d^2z\,e^{-|z|^2} \,D_a(z)\kett{0}_{ab} \nonumber \\& =&
\sqrt{\frac 2\pi}\int_{\mathbb C}
d^2z\,e^{-\frac 32 |z|^2} \sum_{n,m=0}^{\infty}
\frac{1}{n!\,m!} \times\nonumber \\& & 
z^n\,(-z^*)^m\,a^{\dag n}a^n 
\kett{0}_{ab} \nonumber \\&=&
\sqrt{2\pi}\,\frac 23 \sum_{n=0}^{\infty}
\frac{1}{n!} \left(-\frac 23\right)^n\,a^{\dag n}a^n 
\kett{0}_{ab} \nonumber \\&=&
\sqrt{2\pi}\,\frac 23 \sum_{n=0}^{\infty}\left(-\frac 23\right)^n\,
\frac{1}{n!}\,\frac {(a^{\dag }a)!}{(a^{\dag }a-n)!} 
\kett{0}_{ab} \nonumber \\&=&
\sqrt{2\pi}\,\frac 23 \left(\frac 13\right)^{a^\dag a}
\kett{0}_{ab} \nonumber \\&=&
\frac{2\sqrt2}{3}\sum_{n=0}^{\infty}\left(-\frac
13\right)^n \ket{n}_a\otimes\ket{n}_b\nonumber \\&= & 
e^{\scriptsize{\hbox{atanh}}\frac 13(ab-a^\dag b^\dag )}\ket{0} 
\;.\label{twb}
\end{eqnarray}
One recognizes in Eq. (\ref{twb}) the twin-beam state at the output
of a NOPA with total number of photons $N=\bra{\chi}a^\dag a+ b^\dag b
\ket{\chi}=1/4$, corresponding to a gain $G=9/8 $ \cite{dds}.  
More generally, notice that
\begin{eqnarray}
&&\sqrt{\frac {2}{\pi\Delta ^2}}\int_{\mathbb C}
d^2z\,e^{-\Delta^2|z|^2} \,\kett{z}
\nonumber \\&=&\sqrt{1-\lambda ^2}\sum_{n=0}^{\infty}
(-\lambda )^n \ket{n}\otimes\ket{n} 
\nonumber \\&= & 
e^{\scriptsize{\hbox{atanh}}\lambda (ab-a^\dag b^\dag )}\ket{0} 
\;, 
\end{eqnarray}
with $\lambda =(\Delta^2 -1/2)/(\Delta^2+1/2)$.  \par Now let us
evaluate the entangled state $\varrho $ at the output of the cloning
machine. After tracing over the ancillary mode $b$, one has
\begin{eqnarray}
\varrho&=&
\hbox{Tr}_{b}[\proj{{\phi _{out}}}] \nonumber \\&=&
\frac 12 P_{c,a}(\projc {\varphi}\otimes \openone _a) P_{c,a}
\;,\label{trc}
\end{eqnarray}
where $P_{c,a}$ is the projector given by 
\begin{eqnarray}
P_{c,a}&=&
\int_{\mathbb C}d^2z {2\over \pi}
\,e^{-|z|^2}\,D_c^{\dag }(z)\otimes D_a(z) \nonumber \\&
=& V
\left
(\int_{\mathbb C}\frac{d^2z}{\pi}
\,e^{-\frac 12 \,|z|^2}\,D_c^{\dag }(z)\otimes {\openone }_a\right) 
V^\dag \nonumber \\&= & 
V
\left
(\int_{\mathbb C}\frac{d^2z}{\pi}
\,e^{-zc^\dag}\,e^{z^*c}\otimes {\openone }_a\right) 
V^\dag \nonumber \\&= & 
V (\projc{0}\otimes {\openone }_a) V^\dag \;,\label{vv}
\end{eqnarray}
with $V=\exp [\frac \pi 4 (c^\dag a-c a^\dag )]$ that realizes the
unitary transformation 
\begin{eqnarray}
V
\left( \begin{array}{c}c\\ a\end{array}\right)
V^\dag ={1\over \sqrt 2}
\left( \begin{array}{lr} 
1 & -1\\ 
1& 1\end{array}\right) 
\ \left( \begin{array}{c}c\\ a\end{array}\right)
\;.\label{matrix}
\end{eqnarray}
In the last line of Eq. (\ref{vv}) a derivation similar to
Eq. (\ref{twb}) has been followed.  Measuring the quadratures $X_c$
and $Y_a$ over the two clones is then equivalent to perform a measurement
on the original state $\ket{\varphi }_c$, with the measurement
described by the following POVM
\begin{eqnarray}
F(x,y)&=&\frac 12 \hbox{Tr} _{a}[P_{c,a}
\projc{x} \otimes \proja{y} P_{c,a}]
\;,\label{pomxy} 
\end{eqnarray}
where $\ket{x}_c$ and $\ket{y}_a$ denote the eigenstates of  $X_c$ and
$Y_a$, respectively. 
From the following relations \cite{zeta2}
\begin{eqnarray}
&&V^\dag \projc{x} \otimes \proja{y} V \nonumber \\&&= 
2\kett{\sqrt 2 (x-iy)}_{ca}\,{}_{ca} \braa{\sqrt2 (x-iy)}
\;,\label{1}\\& & 
{}_c \langle 0 \kett{z}_{ca}=\frac {1}{\sqrt \pi}
\ket{z^*}_a\;,\label{co1}
\\& & 
V \ket{\alpha }_c\otimes \ket{\beta }_a =
\ket{(\alpha +\beta )/\sqrt 2 }_c\otimes
\ket{(\beta -\alpha )/\sqrt 2}_a \;,\label{co2} 
\end{eqnarray}
[in Eqs. (\ref{co1}) and (\ref{co2}) 
the single-mode states denote coherent states] one obtains
\begin{eqnarray}
F(x,y)=\frac 1 \pi \projc{x+iy}\;,\label{coh}
\end{eqnarray}
namely the coherent-state POVM, which is the well-known optimal POVM
for the joint
measurement of the conjugated quadratures $X_c$ and $Y_c$.  
In fact, from Eqs. (\ref{fout}), (\ref{cl1}) and (\ref{after}), 
one has the following relations between the quantum expectation
values $ \langle \phi _{out}|\cdots 
|\phi _{out}\rangle $ over the output state $|\phi_{out} \rangle $ of
Eq. (\ref{fout}) 
with respect to the values $ \langle \varphi |\cdots 
|\varphi \rangle $ over the original input
state 
\begin{eqnarray}
&&\langle \phi _{out}|g(c,c^\dag)|\phi _{out}\rangle =
\hbox{Tr}_c [\varrho _c\,g(c,c^\dag)] 
\nonumber \\& =&
\int_{\mathbb C}d^2z \,|f(z,z^*)|^2 
\,\langle \varphi |g(c-z,c^\dag-z^*)|\varphi \rangle 
\;, 
\end{eqnarray}
\begin{eqnarray}
&&\langle \phi _{out}|g(a,a^\dag)|\phi _{out}\rangle =
\hbox{Tr}_a [\varrho _a\,g(a,a^\dag)] 
\nonumber \\& =&
\int_{\mathbb C}d^2z \,|\stackrel {\sim }{f}(z,z^*)|^2 
\,\langle \varphi |g(c-z,c^\dag-z^*)|\varphi \rangle 
\;, 
\end{eqnarray}
which holds for any function $g$. 
In particular, for $f(z,z^*)$ given by Eq. (\ref{fz}), one has 
\begin{eqnarray}
&&\langle \phi _{out}|
\Delta X_c^2|\phi _{out}\rangle =
\langle \varphi |
\Delta X_c^2|\varphi \rangle  +\frac 14\;,\label{ad1}  \\& & 
\langle \phi _{out}|\Delta Y_a^2|\phi _{out}\rangle =
\langle \varphi |\Delta Y_c^2|\varphi \rangle +\frac 14\;,\label{ad2}
\end{eqnarray}
namely one achieves the simultaneous measurement
of conjugated quadratures over the input state with minimum added
noise \cite{Yu}, thus proving the optimality of the joint measurement. 
\par The condition in order to obtain identical clones 
$f(z,z^*)=\stackrel{\sim }{f}(z,z^*)$ can be
satisfied also by a bivariate Gaussian of the form
\begin{eqnarray}
f(z,z^*)=\sqrt{ 2\over \pi}\,\exp\left(-\frac{\hbox{Re}^2z}{\sigma ^2}
-\sigma ^2\,\hbox{Im}^2 z\right)\;.\label{sig}
\end{eqnarray}
In the following we will show that in such case the cloning
trasformation becomes optimal for the joint measurement of
noncommuting quadratures at angles which depend on the parameter
$\sigma $ in Eq. (\ref{sig}). In fact, Eq. (\ref{trc}) is replaced by
\begin{eqnarray}
\varrho=
\frac 12 P_{c,a}(\sigma )(\projc {\varphi}\otimes \openone _a)
P_{c,a}(\sigma )
\;,\label{trcsig}
\end{eqnarray}
where the projector $P_{c,a}(\sigma )$ is evaluated as follows
\begin{eqnarray}
&&\!\!\!\!\!\!P_{c,a}(\sigma )=
\int_{\mathbb C}d^2z {2\over \pi}
\,\exp\left(-\frac{\hbox{Re}^2z}{\sigma ^2}
-\sigma ^2\,\hbox{Im}^2 z\right) \times
\nonumber \\& & 
D_c^{\dag }(z)\otimes D_a(z) \nonumber \\&
=& V
\left
[\int_{\mathbb C}\frac{d^2z}{\pi}
\,
\exp\left(-\frac{\hbox{Re}^2z}{2\sigma ^2}
-\frac{\sigma ^2\,\hbox{Im}^2 z}{2}\right)\,
\,D_c^{\dag }(z)\otimes {\openone }_a\right]
V^\dag \nonumber \\&= & 
V\,S_c(\ln \sigma )\,
\left
(\int_{\mathbb C}\frac{d^2z}{\pi}
\,e^{-\frac 12 \,|z|^2}\,D_c^{\dag }(z)\otimes {\openone }_a\right) 
\,S_c^\dag (\ln \sigma )\,V^\dag \nonumber \\&= & 
V \,S_c(\ln \sigma )\,(\projc{0}\otimes {\openone }_a)\,S_c^\dag (\ln
\sigma )\, V^\dag 
\nonumber \\&= & 
S_c(\ln
\sigma )\otimes S_a (\ln
\sigma )\,V (\projc{0}\otimes {\openone }_a) V^\dag \times 
\nonumber \\& &  
S_c^\dag (\ln
\sigma ) \otimes \,S_a^\dag (\ln
\sigma )  \nonumber \\&= &
S_c(\ln
\sigma )\otimes S_a (\ln
\sigma )\,P_{c,a}\,
S_c^\dag (\ln
\sigma ) \otimes \,S_a^\dag (\ln
\sigma )  
\;,\label{vv2}
\end{eqnarray}
with $S_d(r)=\exp[r(d^{\dag 2}-d^2)/2]$ denoting the squeezing operator
for mode $d$ that realizes the unitary transformation
\begin{eqnarray}
S_d^\dag (r) \,d \,S_d(r)=(\cosh r )\,d+(\sinh r )\,d^\dag
\;.\label{scr}
\end{eqnarray}
As in Eq. (\ref{pomxy}), one can evaluate the POVM that is obtained
upon measuring the quadratures $X_c$ and $Y_a$ over the clones. 
From the relations for the quadrature projectors
\begin{eqnarray}
&&S_c^\dag (\ln \sigma )\,\projc{x}
\, S_c(\ln \sigma )=\frac 1 \sigma \projc{x/\sigma }
\;,\nonumber \\& & 
S_a^\dag (\ln \sigma )\,\proja{y}
\, S_a(\ln \sigma )=\sigma \projc{x \sigma }
\;,
\end{eqnarray}
and Eqs. (\ref{1}), (\ref{co1}) and (\ref{co2}), one has 
\begin{eqnarray}
&&F_\sigma (x,y)=\frac 12 \hbox{Tr} _{a}[P_{c,a}(\sigma )
\projc{x} \otimes \proja{y} P_{c,a}(\sigma )]\nonumber \\&= &
\frac 1\pi S_c(\ln \sigma )\,
\left|\frac x \sigma +i \sigma  y\right \rangle _c \,
{\frac{}{}}_c\left \langle \frac x \sigma +i \sigma  y \right |\,S^\dag _c(\ln
\sigma )\label{sque} \\& =&
\frac 1\pi 
D_c(x+iy)\,S_c(\ln \sigma )\projc{0}S^\dag _c(\ln \sigma )\,D^\dag _c (x+iy)
\;.\nonumber
\end{eqnarray}
Eq. (\ref{sque}) shows that the POVM is formally a squeezed 
state. Such kind of POVM is optimal \cite{Yu} 
for the joint measurement 
of the two noncommuting quadrature operators $X_\phi, X_{-\phi}$, 
with $\phi=\hbox{arctg}(\sigma ^2)$. In fact, one has the relations
\begin{eqnarray}
&&\int dx\int dy\, (x\cos \phi \pm y\sin \phi)\,F_\sigma (x,y)=
X_{\pm \phi}\;, \\
&&\int dx\int dy\, (x\cos \phi \pm y\sin \phi)^2\,F_\sigma (x,y)
\nonumber \\&=& X^2_{\pm \phi}+\frac 14 \left|\,\sin(2\phi)\,\right|=
X^2_{\pm \phi}+\frac 12 \left|\,[X_\phi,X_{-\phi}]\,\right|\;,
\end{eqnarray}
namely the outcomes $x\cos \phi \pm y\sin \phi$ trace the
expectation values of the observables $X_{\pm \phi}$ respectively, 
with minimum added noise \cite{Yu}.
\section{Regularization of the universally covariant cloning}\label{reg}
In this section we give a procedure to extend the completely positive
(CP) map for the universal cloning of Werner's paper \cite{Wer} in the case of
infinite dimensional Hilbert space. The procedure is based on a
suitable regularization in order to achieve a trace-preserving map.
In particular, we will show that the universal $1\rightarrow 2$ cloning does not
provide a tool to obtain the joint measurement of noncommuting
observables. Hence, we prove that Werner-type cloning and the cloning
of Ref. \cite{cerf} used in the previous section are different, and they are 
optimal for different purposes.
\par We rewrite here the CP map for $N\rightarrow M$
cloning given in Ref. \cite{Wer}
\begin{eqnarray}
T(\varrho)=\frac{d[N]}{d[M]}\,S_M\,(\varrho \otimes \openone
^{\otimes(M-N)})\,S_M\;,\label{cpc}
\end{eqnarray}
where $d[N]={d+N-1 \choose N}$, $d$ being the dimension of a
single-copy Hilbert space; $S_M$ is the projector on the symmetric
subspace, as mentioned in Sect. II; 
and $\varrho =\ket{\psi}\bra{\psi}^{\otimes N} $ is the initial state of
$N$ identical copies in the state $\ket{\psi}\bra{\psi}$. 
The projector $S_M$ 
can be written in terms of two-site permutation operators $\Pi _{(ij)}$
(transposition), by using recursively the relation \cite{mess}
\begin{eqnarray}
S_M=\frac 1M \left(\openone + \sum_{i=1}^{M-1}\Pi _{(iM)}
\right)\,S_{M-1}\;.
\label{rec}
\end{eqnarray}
The permutation operator $\Pi _{(ij)}$ can be expressed one the Hilbert
space ${\cal H}_i\otimes {\cal H}_j$ as follows\cite{bellpp}
\begin{eqnarray}
\Pi _{(ij)}=\sum_n A_n\otimes A_n^\dag \;, 
\end{eqnarray}
where $\{A_n\}$ are a generic set of operators satisfying the
completeness relation 
\begin{eqnarray}
B=\sum_n \hbox{Tr}[A_n^\dag B ]A_n\;. 
\end{eqnarray}
For example, in the case of $1\rightarrow 2$ cloning for spin $1/2$ one has 
\begin{eqnarray}
S_2&=&\frac 12 (\openone\otimes\openone+ \frac 12\sum_{i=0}^3 \sigma
_i\otimes \sigma _i)\nonumber \\&= & 
\frac 34 \,\openone\otimes\openone+ \frac 14\sum_{i=1}^3 \sigma
_i\otimes \sigma _i
\; 
\end{eqnarray}
where $\sigma _0=\openone $ and $\sigma _i$ ($=1,2,3$) are the 
customary Pauli matrices. 
\par The map in 
Eq. (\ref{cpc}) can be formally extended to infinite dimensional
Hilbert space upon using the transposition operator 
\begin{eqnarray}
\tilde \Pi _{(ij)}=\int \frac{d^2 \alpha}{\pi} 
\,D_i(\alpha )\otimes D^\dag _j(\alpha )\;,\label{tra}
\end{eqnarray}
however the trace-preserving condition on physical CP maps imposes to
replace the identity operator in Eq. (\ref{cpc}) with a normalizable
state. Here we suggest a regularization of $1\rightarrow 2$ cloning in ${\cal
H}_c\otimes {\cal H}_a$ by using Eq. (\ref{tra}) along with the
normalizable (thermal) state $\lambda ^{a\dag a}$, and then we write 
\begin{eqnarray}
\tilde T
(\varrho)&=&K\,\tilde S
_2\,\left(\varrho \otimes \lambda ^{a^\dag a}\right)
\,\tilde S_2 
\;, \label{nonl}
\end{eqnarray}
where $K$ is a constant and 
\begin{eqnarray}
\tilde S_2= \frac 12 \left( \openone _c\otimes \openone _a +
\tilde \Pi _{(ca)}\right)\;. \label{s2t}
\end{eqnarray}
From the identities
\begin{eqnarray}
&&\hbox{Tr}_a [\tilde \Pi _{(ca)}]=\openone _c\;,\qquad 
\hbox{Tr}_c [\tilde \Pi _{(ca)}]=\openone _a\;,\nonumber \\& & 
\tilde \Pi _{(ca)}\,(A\otimes B)=(B\otimes A)\,\tilde \Pi _{(ca)}\;, 
\end{eqnarray}
and the trace-preserving condition $\hbox{Tr}\,\tilde T
(\varrho)=1 $, one obtains the expression for the factor $K$ 
\begin{eqnarray}
K=
2 \left\{\hbox{Tr}\left[ (\openone +\varrho)\,\lambda ^{c^\dag
c}\right]\right\}^{-1}
\;. 
\end{eqnarray}
Notice that the dependence of $K$ on $\varrho $ makes the
transformation in Eq. (\ref{nonl}) nonlinear, however such a nonlinear 
character is vanishing for $\lambda \rightarrow 1$. 
The regularization indeed consists in taking the limit $\lambda \rightarrow
1$. In this case the one-site restricted density matrix is given by 
\begin{eqnarray}
&&\hbox{Tr}_1\,[\tilde T(\varrho)] =
\hbox{Tr}_2\,[\tilde T(\varrho)] \nonumber \\&&= 
\frac 12 \left(\varrho + \frac{\lambda ^{c^\dag c}}{\hbox{Tr}[\lambda
^{a^\dag a}]}\right)
\;,\quad \lambda \rightarrow 1\;, 
\end{eqnarray}
which generalizes the customary depolarizing Pauli channel to the
infinite dimensional case. 
\par In the following we will show that, differently from the cloning
of section \ref{infc}, our regularization of Werner-type cloning does not
allow one to achieve the optimal joint measurement of conjugated
quadratures. In fact, similarly to Eq. (\ref{pomxy}), one 
can evaluate the POVM that corresponds to
separate quadrature measurements over the two clones as follows
\begin{eqnarray}
G(x,y)=\hbox{Tr}_c[K\,\lambda ^{a^\dag a}\,
\tilde S_2\, \projc{x} \otimes \proja{y} 
\,\tilde S_2] \;. 
\end{eqnarray}
Asymptotically, in the limit $\lambda \rightarrow 1$, one rewrites 
\begin{eqnarray}
&&G(x,y)\simeq \frac{1-\lambda }{2}\times \nonumber \\&&
\left(
{}_a \langle y|\lambda ^{a^\dag a}|y \rangle _a
\,\projc{x}+{}_a \langle x|\lambda ^{a^\dag a}|x \rangle _a
\,\projc{y} +\right. \nonumber \\& &
\left. {}_a \langle x|\lambda ^{a^\dag a}|y \rangle _a
\,|x \rangle _c {}_c \langle y|  
+
{}_a \langle y|\lambda ^{a^\dag a}|x \rangle _a
\,|y \rangle _c {}_c \langle x|  \right)
%\nonumber \\&\simeq &
%\sqrt{\frac {1}{\pi (1- \lambda  )}}\,e^{-(1-\lambda )y^2}\,
%\proja{x}+
\;. 
\end{eqnarray}
Notice that one has
\begin{eqnarray}
&&\int dx\int dy\,x\,G(x,y)= \nonumber \\& & 
\frac 12 
X_c + \frac {1-\lambda }{2}\left(\hbox{Tr} [X_a \lambda ^{a^\dag a}]+
\lambda  ^{c^\dag c} X_c +X_c \lambda ^{c^\dag c} \right) \nonumber
\\&\rightarrow & \frac 12 X_c 
\;\\
&&\int dx\int dy\,x^2\,G(x,y)=\nonumber \\& & \frac 12
X^2_c + \frac {1-\lambda }{2}\left(\hbox{Tr} [X^2_a \lambda ^{a^\dag a}]+
\lambda  ^{c^\dag c} X^2_c +X^2_c \lambda ^{c^\dag c} \right) \nonumber
\\&\rightarrow & \frac 12 X^2_c +
\frac 18 \left(1+\frac {2 \lambda}{1- \lambda }  \right)\;,
\end{eqnarray}
and analogous expressions for integration on $y$.  Hence, 
the average values of the variables $x$ and $y$ provide the expectation
values of the quadratures $X_c$ and $Y_c$ (apart from the scaling
factor $1/2$, similar to the shrinking factor of section \ref{fin}). However, 
one can see that the statistical error for such variables diverges 
for $\lambda \rightarrow 1$ since the the second moment goes to
infinity. 
\par The symmetrizer in Eq. (\ref{s2t}) can be rewritten as follows
\begin{eqnarray}
\tilde S_2 &=&\frac 12\, V \,[\openone _c \otimes\openone _a + \int
\frac{d^2 \alpha }{\pi}\,D_c(\sqrt 2\,\alpha ) \otimes \openone _a]
\,V^\dag \nonumber \\&= & \frac 12 \,V \,[\openone _c \otimes\openone _a +
(-)^{c^\dag c}\otimes \openone_a ]\,V^\dag \nonumber \\&=
& V \,\left[\sum_{n=0}^{\infty} |2n \rangle _c {}_c \langle 2n|
\otimes\openone _a \right] \,V^\dag
\;.
\end{eqnarray}
This expression can be more easily compared with the
projector of  Eq. (\ref{vv}) that achieves the cloning transformation
for the optimal joint measurement. 
The different action of the two projectors is clear on the basis of
coherent states. One has
\begin{eqnarray}
&&\tilde S_2 |\alpha \rangle _c |\beta \rangle _a \propto 
|\alpha \rangle _c |\beta \rangle _a +
|\beta \rangle _c |\alpha  \rangle _a \nonumber \\& & 
P_{c,a}|\alpha \rangle _c |\beta \rangle _a \propto 
|(\alpha +\beta )/  2
\rangle _c | (\alpha +\beta )/ 2 \rangle _a 
\;,
\end{eqnarray}
hence the operator $P_{c,a}$ indeed projects on a space that is  
smaller than the symmetric subspace. In fact the cloning map 
${\cal T}(\varrho )=
\frac 12 P_{ca}(\sigma )(\varrho \otimes \openone _a )
P_{ca}(\sigma )$ is not universally covariant, but is covariant only under the 
group of  unitary displacement operators, namely 
\begin{eqnarray}
{\cal T}\left (D(\alpha )\,\varrho \,D^\dag (\alpha )\right)=
D(\alpha )^{\otimes 2}\, {\cal T}(\varrho)\, D^\dag(\alpha )^{\otimes 2} 
\;.
\end{eqnarray}
   
\section{Conclusions}
In this paper we have investigated the possibility of achieving 
joint measurements of noncommuting observables on a single quantum
system by means of quantum cloning.  

We have shown that the universally covariant cloning is not optimal  for joint
measurements, and a suitable non covariant cloning is needed.
Different measures of quality should be used for quantum cloning,
depending on what final use is to be made of the output copies. This
is also indicated by recent studies of different copying machines for
information transfer \cite{munr}. If we want to use quantum cloning to
realise joint measurements, we need to optimise it for a suitable reduced
covariance group, depending on the kind of the desired joint
measurement. For spin 1/2---a finite dimensional
example---the universal cloning optimised by imposing total covariance
\cite{gm,Wer} exhibit added noise in the joint measurement of the spin
components. This shows that for finite dimensional systems the
completely covariant cloning is not optimised for joint measurements,
but in order to achieve optimal joint measurements the cloning
transformations should be optimised with some {\em ad hoc}
procedure. Also in the infinite dimensional case, the suitably
regularized universal covariant cloning map does not allow to achieve
the ideal joint measurement of noncommuting observables.

A restriction of the covariance group in general leads to a higher
fidelity of the cloning transformation, as in the case of
phase covariant cloning \cite{pcc} or for the  cloning map of 
Ref. \cite{cerf}. The last case indeed provides a tool to perform the ideal joint
measurement, as we have shown in section III.

Regarding the experimental feasibility of the schemes of measurement
presented in this paper, we want to stress that a way to implement the
universal cloning transformations \cite{gm} was proposed in \cite{zeil},
with clones as indistinguishable photons, and the final
measurement of the three spin components on the three output copies 
would correspond to nonlinear observables of
radiation, whose measurement is not currently feasible. On the
contrary, the infinite dimensional case is much more realistic, since
the $1\to 2 $ cloning transformation considered in section III can be
achieved experimentally by means of a sequence of parametric
amplifiers \cite{dds}, and the quadrature measurements are obtained by
customary homodyne detectors.

This work was supported in part by the European Union project 
EQUIP (contract IST-1999-11053) and by
Ministero dell'Universit\`a e della Ricerca Scientifica e Tecnologica
under the project ``Quantum information transmission and processing:
quantum teleportation and error correction''. One of us (M.F.S.) is also 
acknowledging support from ESF-QIT program.

\end{document}